\documentstyle[aps,prl,twocolumn,epsfig]{revtex}

\title{Vortex formation in a stirred Bose-Einstein condensate}
\author{K. W. Madison, F. Chevy, W. Wohlleben$^{\dagger}$, J. Dalibard}
\address{Laboratoire Kastler Brossel$^*$, D\'epartement de Physique de
l'Ecole Normale Sup\'erieure\\
24 rue Lhomond, 75005 Paris, France}
\date{final version: December 22, 1999}
\begin{document}
\maketitle 

\begin{abstract}
Using a focused laser beam we
stir a Bose-Einstein condensate
of $^{87}$Rb confined in a magnetic trap
and observe the formation of a vortex for a
stirring frequency exceeding a critical value.
At larger rotation frequencies we produce states of the
condensate for which up to four vortices are simultaneously present.
We have also measured the lifetime of the single vortex state
after turning off the stirring laser beam.
\end{abstract}

{\pacs 03.75.Fi, 67.40.Db, 32.80.Lg}

\vskip 5mm

Rotations in quantum physics constitute a 
source of counterintuitive predictions and results as 
illustrated by the famous ``rotating bucket'' experiment
with liquid helium.
When an ordinary fluid is placed in a rotating container, 
the steady state corresponds to a rotation of the fluid as a
whole together with the vessel. Superfluidity, first observed in liquid HeII, 
changes dramatically this behavior \cite{Lifshitz,Donnelly}.
For a small enough rotation frequency, no motion 
of the superfluid is observed; while above a critical frequency, lines
of singularity appear in its velocity field.  These singularities,
referred to as vortex filaments, correspond to a quantized circulation
of the velocity ($ nh/m$ where
$n$ is an integer, and $m$ the mass of a particle of the fluid) 
along a closed contour around the vortex.
In this letter we report the observation of such vortices in a stirred
gaseous condensate of atomic rubidium. We determine the critical 
frequency for their formation, and we analyze their metastability 
when the rotation of the confining ``container'' is stopped.

The interest in vortices for gaseous condensates 
is that, due to the very low density, the theory is tractable in these systems
and the diameter of the vortex core, which is on the order of the
healing length, is typically three orders of magnitude larger than
in HeII.  At this scale, further improved by a ballistic expansion, 
the vortex filament is large enough to be observed optically.  
The generation of quantized vortices in gaseous samples has been
the subject of numerous theoretical studies since the first
observations of Bose-Einstein condensation in atomic gases
\cite{Anderson95,Bradley957,Davis95,Fried98}.  Two schemes
have been considered.  The first one uses laser beams
to engineer the phase of the condensate wave function and produce
the desired velocity field \cite{Marzlin,Dum,Petrosyan,Williams,Dobrek99}.
Recently this scheme \cite{Williams} has been successfully applied to a binary mixture
of condensates, resulting in a quantized rotation of one of the
two components around the second one \cite{Matthews99}.
Phase imprinting has also been used for the generation of solitons inside
a condensate \cite{Burger99,Phillips}.

The second scheme, which is explored in the present work, is 
directly analogous to the rotating bucket experiment \cite{Leggett,Stringari99}.
The atoms are confined in a static, cylindrically-symmetric
Ioffe-Pritchard magnetic trap upon which we 
superimpose a non-axisymmetric,
attractive dipole potential created by a stirring laser beam.
The combined potential leads to a cigar-shaped harmonic trap
with a slightly anisotropic transverse profile.  The
transverse anisotropy is rotated at angular frequency $\Omega$ 
as the gas is evaporatively cooled
to Bose-Einstein condensation, and it plays the role of the bucket wall roughness.

In this scheme, the formation of vortices
is --in principle-- a consequence of thermal equilibrium. 
In the frame rotating at the same
frequency as the anisotropy, the Hamiltonian is time-independent
and one can use a standard thermodynamics approach to 
determine the steady-state of the system.
In this frame, the Hamiltonian can be written 
$\tilde H=H-\Omega L_z$, where $H$
is the Hamiltonian in the absence of rotation,
and $L_z$ is the total orbital angular momentum
along the rotation axis. 
Above a critical rotation frequency, $\Omega_{\rm c}$, 
the term $-\Omega L_z$ can favor the creation of a
state where the condensate wave function
has an angular momentum $\hbar$ along the $z$ axis
and therefore contains a vortex filament 
\cite{Baym96,Stringari96,Sinha97,Lundh97,Butts99,Feder99,Fetter98,Castin99,Caradoc}. 
The density of the condensate at the center of the vortex 
is zero, and the radius of the vortex core is of the order of 
the healing length $\xi=(8\pi a \rho)^{-1/2}$,
where $a$ is the scattering length characterizing the 2-body interaction, and
$\rho$ the density of the condensate \cite{Dalfovo99}.

The study of a vortex generated by this second route
allows for the investigation of several debated questions
such as the fate of the system when the rotating velocity 
increases above $\Omega_{\rm c}$. 
This could in principle lead to the formation of a single vortex with $n>1$
at the center of the trap; however, this state has been shown to be either dynamically 
or thermodynamically unstable \cite{Lifshitz,Butts99,Feder99,Fetter98,Castin99,Pu}.
The predicted alternative for large rotation frequencies consists
of a lattice of $n=1$ vortices.
Another important issue is the stability of the current
associated with the vortex once the rotating anisotropy
is removed\cite{Rokhsar97}.
 
Our experimental set-up has been described in detail previously
\cite{Soding99}.  We start with $10^{9}$ $^{87}$Rb atoms 
in a magneto-optical trap
which are precooled and then transferred into an Ioffe-Pritchard 
magnetic trap.  
The evaporation radio frequency starts at $\nu_{\rm rf}=15$~MHz and 
decreases exponentially to $\nu_{\rm rf}^{\rm (final)}$ in 25~s with a time constant 
of 5.9~s. Condensation occurs at 
$\Delta\nu_{\rm rf}=\nu_{\rm rf}^{(\rm final)}-\nu_{\rm rf}^{(\rm min)}\simeq
50$~kHz, with  $2.5\;10^6$ atoms and a temperature $500$~nK. 
Here $\nu_{\rm rf}^{(\rm min)}=430\;(\pm 1)$~kHz is the
radio frequency which empties completely the trap.
The slow oscillation frequency of the 
elongated magnetic trap is $\omega_z/(2\pi)= 11.7$~Hz ($z$ is horizontal in 
our setup), while the transverse oscillation frequency is $\omega_\bot/(2\pi)=
219$~Hz.  For a quasi-pure condensate with $10^5$ atoms,
using the Thomas-Fermi approximation,
we find  for the radial and longitudinal sizes 
of the condensate $\Delta_{\bot}=2.6\,\mu$m and $\Delta_{z}=49\,\mu$m, respectively. 

When the evaporation radio frequency $\nu_{\rm rf}$ reaches the value $\nu_{\rm rf}^{\rm{(min)}}+80$~kHz, 
 we switch on the stirring laser beam which  
propagates along the slow axis of the magnetic trap.
The beam waist is $w_s=20.0\; (\pm\,1)$~$\mu$m and 
the laser power $P$ is 0.4~mW.
The recoil heating induced by this far-detuned beam (wavelength $852$~nm)
is negligible.  Two crossed acousto-optic modulators, combined with 
a proper imaging system, then allow for an arbitrary
translation of the laser beam axis with respect to the symmetry axis 
of the condensate.

The motion of the stirring beam
consists in the superposition of a fast and a slow component.
The optical spoon's axis is toggled 
at a high frequency (100~kHz) between two symmetric
positions about the trap axis $z$.
The intersections of the stirring beam axis and the $z=0$ plane are 
$\pm a (\cos\theta\, {\bf u}_x+\sin \theta \,{\bf u}_y)$,
where the distance $a$ is $8\;\mu$m.
The fast toggle frequency is chosen to be much larger than
the magnetic trap frequencies so that the atoms experience
an effective two-beam, time averaged potential. 
The slow component of the motion is a uniform rotation of the angle 
$\theta=\Omega t$. The value of the angular frequency $\Omega$ is 
maintained fixed during the evaporation at a value chosen between 0 
and 250~rad~${\rm s}^{-1}$.

Since $w_s \gg \Delta_{\bot}$,
the dipole potential, proportional to the power of the stirring beam, 
is well approximated by $m\omega_\bot^2 (\epsilon_X X^2 +\epsilon_Y Y^2)/2$.
The $X,Y$ basis is rotated with respect to the fixed axes ($x,y$) by the
angle $\theta(t)$, and $\epsilon_X=0.03$ and $\epsilon_Y=0.09$ for the 
parameters given above \cite{verif}.  
The action of this beam is essentially a slight modification of 
the transverse frequencies of the magnetic trap
while the longitudinal frequency is nearly unchanged.  
The overall stability of the stirring beam 
on the condensate appears to be a crucial element for the 
success of the experiment, and we estimate that 
our stirring beam axis is fixed to and stable on the condensate 
axis to within 2~$\mu$m. We checked that for $\Omega < \Omega_{\rm c}$ the stirring beam does not
affect the evaporation.  

For the data presented here, the final frequency
of the evaporation ramp was chosen just above $\nu_{\rm rf}^{(\rm min)}$
($\Delta\nu_{\rm rf} \in [3,6]$~kHz). After the end of the evaporation ramp, we let the system
reach thermal equilibrium in this ``rotating bucket'' for a duration 
$t_{\rm r}=500$~ms in the presence of an rf shield 30~kHz above
$\nu_{\rm rf}^{(\rm final)}$.
The vortices induced in the condensate by the optical spoon
are then studied using a time-of-flight analysis.  
We ramp down the stirring beam slowly (in 8~ms)
to avoid inducing additional excitations in the condensate,
and we then switch off the magnetic field and
allow the droplet to fall for $\tau=27$~ms. 
Due to the atomic mean field energy, the initial cigar shape
of the atomic cloud transforms into
a pancake shape during the free fall. 
The transverse $xy$ and $z$ sizes
grow by a factor of $40$ and $1.2$ respectively \cite{Castin96}. 
In addition, the core size of the vortex should expand at 
least as fast as the transverse size of the condensate 
\cite{Castin96,Lundh98,Dalfovo00}.
Therefore a vortex with an initial diameter
$2\xi=0.4\;\mu$m for our experimental parameters is expected to
grow to a size of 16~$\mu$m.

At the end of the time-of-flight period, we illuminate the atomic 
sample with a resonant probe laser for 20~$\mu$s. 
The shadow of the atomic cloud in the probe beam is imaged 
onto a CCD camera with an optical resolution $\sim 7\,\mu$m.
The probe laser propagates along the $z$-axis 
so that the image reveals the column density of the cloud after expansion
along the stirring axis.
The analysis of the images, which proceeds along the same
lines as in \cite{Soding99}, gives access to the number of condensed
$N_0$ and uncondensed $N'$ atoms and to the temperature $T$.
  Actually, for the present data, the uncondensed part
of the atomic cloud is nearly undetectable, and we can only give
an upper bound for the temperature $T<80$~nK.

Figure \ref{photos} shows a series of five pictures 
taken at various rotation frequencies $\Omega$.
They clearly show that for fast enough rotation frequencies, we can generate
one or several (up to 4) ``holes'' in the transverse 
density distribution corresponding to vortices.
We show for the 0- and 1-vortex cases a cross-section of the
column density of the cloud along a transverse axis. 
The 1-vortex state exhibits a spectacular dip at the center
(up to 50\,\% of the maximal column density)
which constitutes an unambiguous signature of the
presence of a vortex filament.  The diameter of the vortex core
following the expansion is measured at the half max of the dip
to be $\sim 20\,\mu$m.

For a systematic study of the vortex stability 
domain, we have varied in steps of 1~Hz the rotation frequency 
for a given atom number and temperature.
For each frequency, we infer from the absorption image
the number of vortices present, and the results are shown
in Fig. \ref{edges}.  Below a certain frequency, 
we always obtain a condensate with no vortices. 
Then, in a zone with a 2~Hz width, we obtain 
condensates showing randomly 0 or 1 vortex. Increasing $\Omega$,
we arrive at a relatively large frequency interval (width 10 Hz)
where we systematically observe a condensate with a single vortex present.
Our value for the critical frequency is notably larger than the predicted value
of 91~Hz \cite{Lundh97} (see also \cite{Baym96,Sinha97,Castin99,Feder99}).
This deviation may be due to the marginality of the Thomas-Fermi approximation
for our relatively low condensate number.
If $\Omega$ is increased past the upper edge of the 1-vortex zone, 
multiple vortices, as shown in Fig. \ref{photos}, are observed .
The range of stability of the multiple vortex zones 
appears to be much smaller than that for the 1-vortex zone. 
The 3-vortex zone, for instance, seems to be stable over only 3-4 Hz
and is complicated by the occasional appearance of a 
2-vortex or a 4-vortex condensate.
At this stage of the experiment, it is difficult to 
determine whether these shot-to-shot fluctuations are due to a 
lack of experimental reproducibility
or to the fact that these various
states all have comparable energies
and therefore all have a reasonable probability to occur
for the range of parameters in question.
Finally, when $\Omega$ is increased past the range of stability for
the multiple vortex configuration, the density profile of the condensate
takes on a turbulent structure, and the condensate completely 
disappears for $\Omega$ larger than 210~Hz, which should be compared with 
the average transverse frequency of the magnetic + laser dipole potential (226~Hz).

It is remarkable that the multiple vortex configurations most often
occur in a symmetric arrangement of the vortex cores: an equilateral
triangle and a square for the 3-vortex and the 4-vortex cases
respectively. This finding supports the theoretical analysis of \cite{Castin99}
which shows that vortices rotating in the same direction experience an
effective repulsive interaction, which in turn favors these stable configurations (see also \cite{Butts99}).

The final question addressed in this letter concerns the lifetime of a vortex state
in an axisymmetric trap.  Without a rotating anisotropy, the vortex state 
is no longer the lowest energy state of the system, and after the anisotropy is
removed, one expects that the gas will eventually relax to a condensate 
with no vortex plus a slightly larger thermal component, bolstered
by the energy contained in the vortex state.
Figure \ref{decay} presents the experimental study of
the single-vortex state lifetime at two different condensate parameters.
We choose a rotation frequency $\Omega$ in the middle of the 1-vortex 
range of stability, and we let the vortex form as before in the presence of the stirring beam.
Then we switch off the stirring beam, and we allow the gas to evolve in the pure
magnetic trap for an adjustable time. Finally, we perform
a time-of-flight analysis to determine whether the vortex is present or not.
Each point in the two lifetime curves represents the 
average of 10 shots,  where we have plotted the fraction of pictures 
showing unambiguously a vortex as a function of time \cite{blind}. 
We deduce from this curve a characteristic lifetime of the 
vortex state in the range 400 to 1000~ms with a 
clear non-exponential decay behavior. 
In addition, we observed that at long times the vortex rarely 
appears well centered as it does immediately after
formation.

To summarize, we have reported the formation of vortices in a 
gaseous Bose-Einstein condensate when it is stirred by a laser beam which produces a
slight rotating anisotropy.  A natural extension of this work is
to study the superfluid aspects of this system \cite{Raman99}
by investigating the dynamics for vortex nucleation and decay.
An important question is the role of the thermal component.
For instance, nucleation can occur either by transfer of angular momentum
from this component to the condensate
or directly from a dynamical instability of the non-vortex state \cite{Feder99,Castin99,Machida99,Svidzinsky98A}.
Also the decay of the single vortex state may be due to the coupling
with a non-rotating thermal component \cite{Fedichev99} or due to the instability induced
by the residual, fixed anisotropy of our magnetic trap (measured to be $\omega_x/\omega_y=1.012 \pm .002$) \cite{rotation}.
Finally, this type of experiment
gives access, in principle, to the elementary excitations of the vortex
filament \cite{Dodd97,Zambelli98,Svidzinsky}, the study of which
might reveal new aspects of the superfluid properties of these systems.

{\acknowledgments
We thank Y. Castin, C. Cohen-Tannoudji, C. Deroulers,
D. Gu\'ery-Odelin, C. Salomon,
G. Shlyapnikov, S. Stringari, and the ENS Laser cooling
group for several helpful discussions and comments. 
This work was partially supported by CNRS, Coll\`{e}ge de France,
DRET, DRED and EC (TMR network ERB FMRX-CT96-0002). This material is
based upon work supported by the North Atlantic Treaty Organization
under an NSF-NATO grant awarded to K.M. in 1999.

$^\dagger$ permanent address: Max Planck Institute f\"ur KernPhysik, Heidelberg, Germany.

$^*$ Unit\'e de Recherche de l'Ecole normale sup\'erieure et de
l'Universit\'e Pierre et Marie Curie, associ\'ee au CNRS.
}

\newpage
\begin{figure}
\caption{Transverse absorption images of a Bose-Einstein condensate
stirred with a laser beam (after a 27~ms time-of-flight).
For all five images, the condensate number is $N_0=(1.4\;\pm 0.5)\;10^{5}$
and the temperature is below 80~nK.
The rotation frequency $\Omega/(2\pi)$ is respectively
(c) 145~Hz; (d) 152~Hz; (e) 169~Hz; (f) 163~Hz; (g) 168~Hz.
In (a) and (b) we plot the variation of the optical thickness
of the cloud along the horizontal transverse axis for the images (c) (0 vortex)
and (d) (1 vortex).}
\label{photos}
\end{figure}

\begin{figure}
\caption{Edges of the regions of stability for the 0, 1, and multiple vortex 
configurations. The condensate number is $N_0=(2.3\;\pm 0.6)\;10^{5}$
and the temperature below 80~nK.}
\label{edges}
\end{figure}

\begin{figure}
\caption{Fraction of images showing a vortex as a function of the
time spent by the gas in the axisymmetric trap after the end of the stirring phase
for two condensate conditions.  The sets of data correspond to condensate numbers of
$N_0^{(\circ)}=(2.3\;\pm 0.6)\;10^{5}$ and $N_0^{(\bullet)}=(1.2\;\pm 0.3)\;10^{5}$
which were obtained with $\Delta\nu_{\rm rf}=$6 and 3 kHz respectively.  
This implies that $T^{(\circ)}>T^{(\bullet)}$, where both temperatures are below
our detection limit of 80~nK.}
  
\label{decay}
\end{figure}


\begin{references}
\bibitem{verif} We observed that 
the vortex formation is not sensitive to the exact value 
of $\epsilon_X,\epsilon_Y$, and a single vortex state under the
same conditions was successfully created with a laser power 
of half and twice the value of 0.4~mW used here.
\bibitem{blind} To eliminate a possible ``psychological bias'', the images were
examined in a random order after collection.
\bibitem{rotation} Our observation that the quantities 
$N/T^3$ at the condensation point
with and without the rotating spoon (160 Hz) were
equal (to within 15\%) indicates that the gas was not yet rotating
at this stage in the evaporation.
\end{references}
\end{document}